\documentclass{emulateapj}

\begin{document}

\shortauthors{Luhman}

\shorttitle{Census of the TW Hya Association}

\title{A Census of the TW Hya Association with Gaia\altaffilmark{1}}

\author{K. L. Luhman\altaffilmark{2,3}}

\altaffiltext{1}
{Based on observations made with the Gaia mission, the Two Micron All
Sky Survey, the Wide-field Infrared Survey Explorer, 
Gemini Observatory, and European Southern Observatory telescopes 
at La Silla and Paranal Observatories.}

\altaffiltext{2}{Department of Astronomy and Astrophysics,
The Pennsylvania State University, University Park, PA 16802, USA;
kll207@psu.edu}

\altaffiltext{3}{Center for Exoplanets and Habitable Worlds, The
Pennsylvania State University, University Park, PA 16802, USA}

\begin{abstract}

I have used high-precision photometry and astrometry from the third 
data release of Gaia to perform a survey for members of the TW Hya 
association (TWA).
I have identified candidate members that appear to share similar kinematics and
ages with bona fide members compiled by \citet{gag17} 
and I have assessed their membership using radial velocities and spectroscopic
diagnostics of age from various sources. 
My new catalog of adopted members contains 67 Gaia sources
in 55 systems.  The histogram of spectral types for TWA peaks near M5 
($\sim0.15$~$M_\odot$), resembling the distributions measured for
other nearby young associations. The $UVW$ velocities of its members
indicate that the association is expanding. The rate of expansion corresponds
to an age of 9.6$^{+0.9}_{-0.8}$~Myr. In a Gaia color-magnitude diagram,
the members of TWA exhibit well-defined sequences of single stars and
unresolved binary stars.
The combined sequence of low-mass stars in TWA is indicative of an age
of $11.4^{+1.3}_{-1.2}$~Myr when compared to the sequence for Upper
Centaurus-Lupus/Lower Centaurus-Crux, for which an age of 20~Myr is assumed.
Based on these expansion and isochronal ages, I have adopted an age of 
10$\pm$2~Myr for TWA.  Finally, I have used mid-infrared photometry from the
Wide-field Infrared Survey Explorer to check for excess emission
from circumstellar disks among the TWA members. Fourteen members have
detected disks, all of which have been reported in previous studies.
The fraction of members at $\leq$M6 ($\gtrsim0.1$~$M_\odot$) that have full,
transitional, or evolved disks is 10/52=$0.19^{+0.08}_{-0.06}$. 
That value is similar to the fraction previously measured for the
Upper Sco association, which is roughly coeval with TWA.

\end{abstract}

\section{Introduction}
\label{sec:intro}

Because of its proximity and youth ($\sim$40--100~pc, 10~Myr), the 
TW~Hya association (TWA) is one of the most important stellar
populations in the solar neighborhood.
TW~Hya was identified as a possible young star based on H$\alpha$ emission 
\citep{hen76} and it was confirmed as such through the detection of Li 
absorption at 6708~\AA\ and additional emission lines \citep{her78,ruc83}.
Candidates for young stars associated with TW~Hya have been
selected via their positions near sources detected
by the Infrared Astronomical Satellite \citep{del89,gre92,zuc93}, 
X-ray emission \citep{kas97,jen98,ste99,web99,zuc01,son03,loo10a},
UV emission \citep{rod11,shk11,bin20}, and optical and infrared (IR) colors
\citep{giz02,sch05,loo07,loo10b,sch12b,sch16,gag14b,gag17,kel15}, often 
in conjunction with proper motions measured from wide-field imaging surveys.
For some of those candidates, membership in TWA has been further constrained 
using measurements of radial velocities 
\citep[e.g.,][]{rei03,tor03,ell14,mal14,kel16,kid19}, parallaxes 
\citep{bil07,giz07,duc08,tei08,duc14,wei13,don16,bes20}, and the moving cluster
distance \citep{mam05}. In one of the more recent membership studies of TWA,
\citet{gag17} classified 24 and 11 systems as bona fide members and
high-likelihood candidates, respectively.

The identification of members of nearby associations has been
greatly facilitated by the Gaia mission \citep{per01,deb12,gaia16b}, 
which is measuring high-precision photometry, proper motions, and
parallaxes for more than a billion stars down to $G\sim20$, corresponding 
to masses of $\sim0.01$--0.015~$M_\odot$ in TWA \citep{bar15,cha23}.
\citet{gag18d} used the second data release of Gaia (DR2) to identify candidate
members of 27 nearby associations, including nine candidates for TWA, most 
of which have lacked the spectroscopy needed for confirmation of membership.

In this paper, I use the third data release of Gaia \citep[DR3,][]{bro21,val22}
to identify candidate members of TWA and I assess their membership
with spectra and radial velocity measurements from various sources. 
I then use my catalog of adopted members to study the association in terms 
of its initial mass function (IMF), kinematic and isochronal ages, and 
circumstellar disks.

\section{Search for New Members of TWA}

\subsection{Identification of Candidate Members}
\label{sec:ident}

For my survey of TWA, I have used the following data from Gaia DR3:
photometry in bands at 3300--10500~\AA\ ($G$), 3300--6800~\AA\ ($G_{\rm BP}$),
and 6300-10500~\AA\ ($G_{\rm RP}$); proper motions and parallaxes
($G\lesssim20$); and radial velocities ($G\lesssim15$).
For parallactic distances, I adopt the geometric values estimated by
\citet{bai21} from the DR3 parallaxes. 
In addition, I have made use of photometry in three near-IR
bands ($JHK_s$) from the Point Source Catalog of the Two Micron All Sky Survey
\citep[2MASS,][]{skr03,skr06} and photometry in four mid-IR bands
from the AllWISE Source Catalog of the Wide-field Infrared Survey Explorer 
\citep[WISE,][]{wri10,wri13,cut13a}.
The WISE bands are centered at 3.4, 4.6, 12, and 22~$\mu$m,
which are denoted as W1, W2, W3, and W4, respectively.
For members of TWA, 2MASS is slightly deeper than Gaia, and WISE is deeper
than both 2MASS and Gaia. 

As done in my previous surveys of young associations 
\citep[e.g.,][]{esp17,luh20u,luh22sc}, I have analyzed
the Gaia astrometry in terms of a ``proper motion offset"
($\Delta\mu_{\alpha,\delta}$), which is defined as the difference between the
observed proper motion of a star and the motion expected at the celestial
coordinates and parallactic distance of the star for a specified
space velocity. The use of this metric minimizes projection effects,
which is important for associations like TWA that cover a large area of sky.
For my survey of TWA, the proper motion offsets are calculated relative to the
motions expected for a velocity of $U, V, W = -12, -18, -6$~km~s$^{-1}$,
which approximates the median velocity of TWA members 
\citep[][Section~\ref{sec:uvw}]{gag17}. 

To perform a census of TWA, I began by considering the most probable
members from previous work. In a detailed study of TWA, \citet{gag17}
compiled a sample of 30 objects in 24 systems that they classified as
bona fide members of TWA. While presenting an update to their algorithm
for identifying members of young associations, \citet{gag18b} also
included a list of bona fide TWA members, which was the same as in
\citet{gag17} except for the exclusion of TWA~9 by the newer study.
I find that the data for TWA~9 are consistent with those of the other bona
fide members, so I have used the sample from \citet{gag17} to guide my survey.

\begin{figure}
\epsscale{1.2}
\plotone{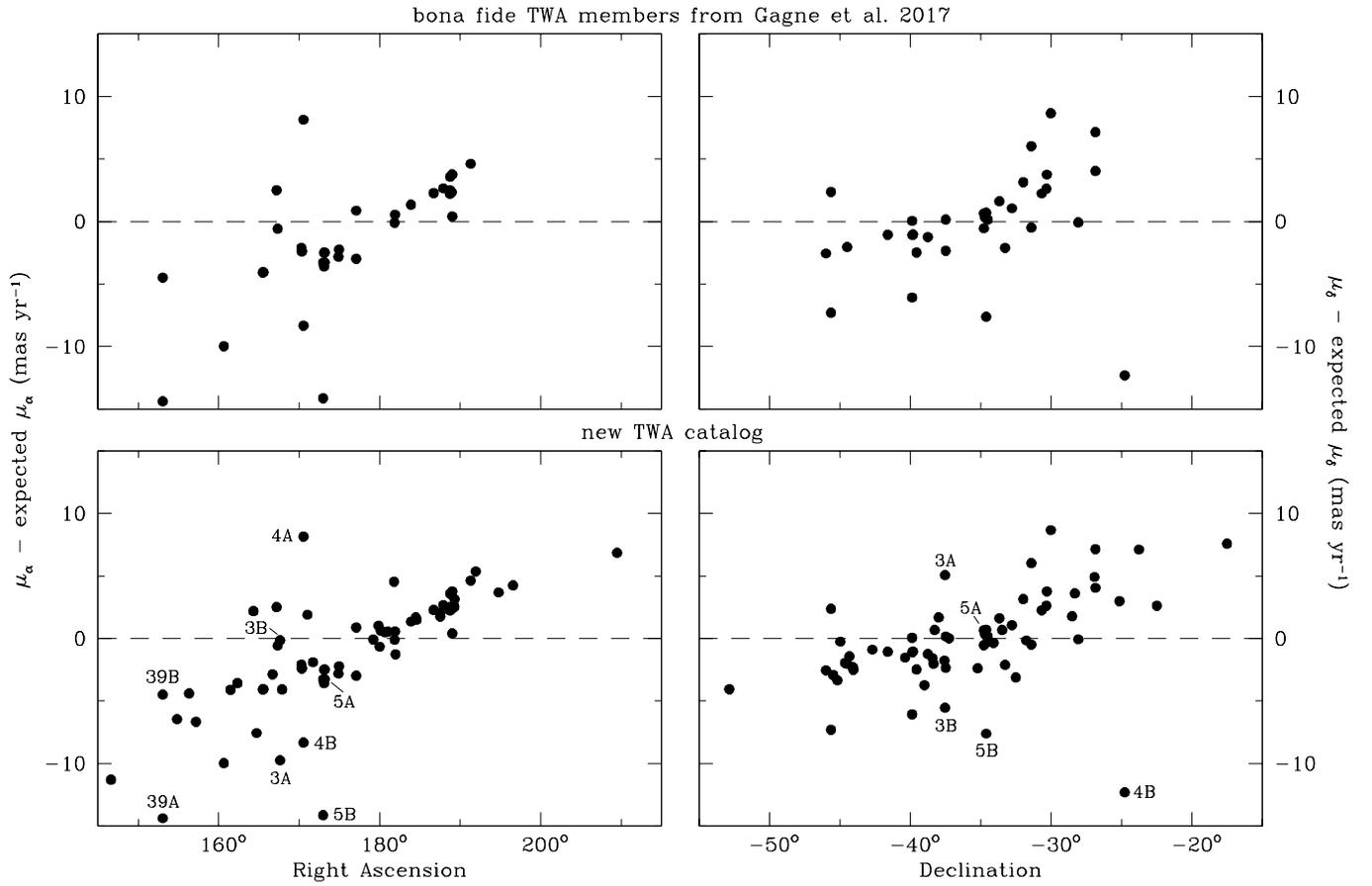}
\caption{
Proper motion offsets versus equatorial coordinates 
for the bona fide members of TWA compiled by \citet{gag17} (top)
and the adopted members from this work (bottom).
Outliers are labeled with their TWA numbers. TWA~4A is beyond the boundaries
of the diagrams on the right.}
\label{fig:pp}
\end{figure}

\begin{figure}
\epsscale{1.2}
\plotone{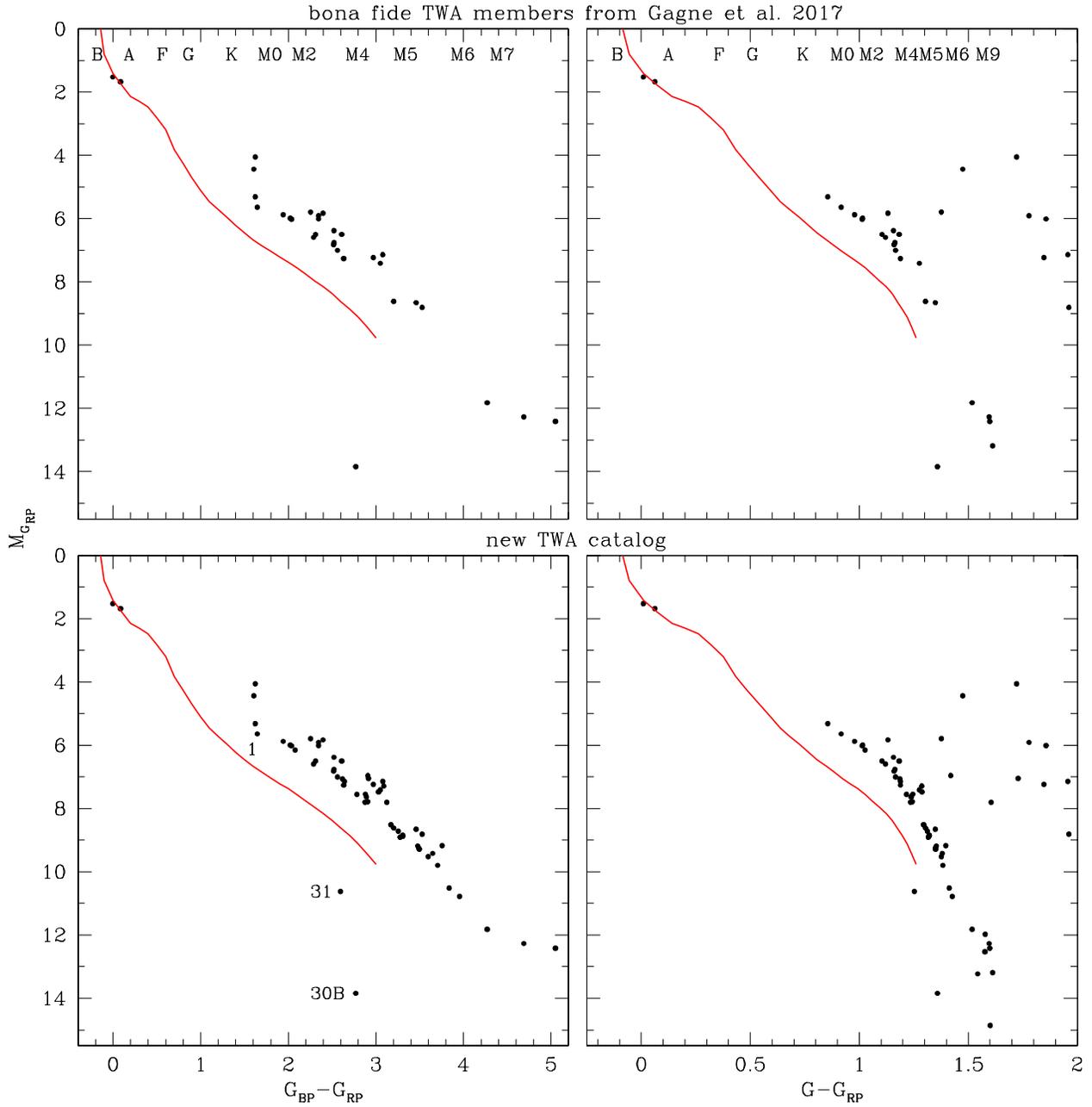}
\caption{
$M_{G_{\rm RP}}$ versus $G_{\rm BP}-G_{\rm RP}$ and $G-G_{\rm RP}$
for the bona fide members of TWA compiled by \citet{gag17} (top) 
and the adopted members from this work (bottom). 
Three stars that appear below the sequence in $G_{\rm BP}-G_{\rm RP}$ 
are labeled with their TWA numbers.
Each CMD includes a fit to the single-star sequence of the
Pleiades \citep[red line,][]{luh23}. The spectral types that correspond 
to the colors of young stars are indicated in the top CMDs \citep{luh22sc}.
}
\label{fig:cmd1}
\end{figure}

In the top row of Figure~\ref{fig:pp}, I have plotted the bona fide TWA
members from \citet{gag17} in diagrams of $\Delta\mu_{\alpha}$ versus right
ascension and $\Delta\mu_{\delta}$ versus declination.
A correlation is present in each diagram, which is suggestive of expansion.
The expansion of TWA is better characterized through analysis of $UVW$
velocities in Sections~\ref{sec:uvw} and \ref{sec:uvw2}.  The outliers among 
the bona fide members in Figure~\ref{fig:pp} are components of binary systems.
Among those outliers, the components of a given binary
tend to straddle the bulk of the bona fide members (i.e., the average
proper motion offsets of a binary's components are consistent with
the overall population).
In the top row of Figure~\ref{fig:cmd1}, I show the bona fide TWA members
in color-magnitude diagrams (CMDs) consisting of $M_{G_{\rm RP}}$ versus
$G_{\rm BP}-G_{\rm RP}$ and $G-G_{\rm RP}$. I have excluded photometry with 
errors greater than 0.1~mag. The errors for TWA~30A exceed that threshold in 
$G_{\rm BP}$ and $G_{\rm RP}$, so it does not appear in either CMD.
For comparison, I have included in each CMD a fit to the single-star sequence
of the Pleiades cluster \citep[$\sim120$~Myr,][]{sta98,dah15} from
\citet{luh23}.
TW~Hya (TWA~1) is slightly fainter than the sequence formed by other members
in the $G_{\rm BP}-G_{\rm RP}$ CMD while TWA~30B is well below the sequence
in both CMDs. The former is likely caused by UV excess emission related
to accretion \citep{ruc83,muz00} and the latter may be evidence of an
edge-on disk \citep{loo10b}.

To identify candidate members of TWA, I selected sources from Gaia DR3 that 
(1) are located within a spatial volume extending well beyond the bona fide
members but that do not overlap with candidate members of Upper 
Centaurus-Lupus/Lower Centaurus-Crux (UCL/LCC) from
\citet{luh22sc} (see Section~\ref{sec:uvw}),
(2) have proper motion offsets that are similar to those of the bona fide
members or that are consistent with their pattern of expansion when
extended in equatorial coordinates, and
(3) appear near or above the lower envelope of the sequence of bona fide
members in at least one CMD and do not appear below the sequence in either CMD.
Since stars with disks can appear underluminous in CMDs, as mentioned earlier
for TWA~1 and TWA~30B,
I also considered Gaia sources that satisfied the first two criteria, 
appeared below the sequence in a CMD, and exhibited mid-IR excess emission in
data from WISE. For the resulting candidates, I searched data archives
and previous studies for spectroscopic measurements of age diagnostics 
(e.g., Li) and radial velocities. When spectroscopy was unavailable,
I pursued new observations to measure spectral types and age diagnostics
(Section~\ref{sec:spec}). I rejected candidates if spectroscopy indicated ages
older than those of the bona fide members 
or if radial velocity measurements and Gaia astrometry produced $UVW$ 
velocities that are inconsistent with membership (Section~\ref{sec:uvw}).
In the end, I arrived at a catalog of 67 adopted TWA members that have entries 
in Gaia DR3, which consist of 35 Gaia sources in 24 systems that correspond 
to the bona fide members from \citet{gag17} and 32 additional Gaia sources
in 31 systems. One of these sources, TWA~32B, has an entry in Gaia DR3 but 
lacks a parallax measurement. The companions TWA~2B and TWA~27B are absent
from my catalog since they lack Gaia detections.
Gaia usually detects both components of a pair like TWA~2A and 2B,
which has a separation of $0\farcs56$ and $\Delta K=0.8$ \citep{web99}.
TWA~27B has a wider separation of $0\farcs78$ from its primary \citep{cha04},
which is a brown dwarf \citep{giz02}, but a detection by Gaia is not expected 
given its very low luminosity.

To show the proper motion offsets and CMDs for my full catalog of adopted 
members, I have included them in the bottom rows of Figures~\ref{fig:pp} and
\ref{fig:cmd1}. In the $G_{\rm BP}-G_{\rm RP}$ CMD, the TWA members exhibit
a narrow, well-defined lower sequence that likely consists of single stars.
An elevated sequence is evident as well, which likely
corresponds to unresolved binaries. To my knowledge, (1) these CMDs are
the first to resolve the single- and double-star sequences in TWA and 
(2) TWA is one of the youngest populations in which that has been done.
The sequences of young stellar populations in CMDs and other versions of
Hertzsprung-Russell (H-R) diagrams are broadest in star-forming regions and
become more narrow at older ages, likely because of decreasing variability 
from spots and accretion, the removal of most extinction as the natal
cloud disperses, and the fact that a given age spread
corresponds to a smaller range of luminosities at older ages.
As a result, detecting separate sequences for single and binary
stars is more easily done with older populations. Doing so is also facilitated
by accurate photometry, precise relative distances for the members
of a population, and low contamination from nonmembers.
Relative to most other clusters and associations, TWA has more accurate
photometry and distances because of its proximity and less (negligible)
variable extinction because of its location within the Local Bubble.
In comparison, the single- and double-star sequences for Upper Sco and UCL/LCC
(10 and 20~Myr) are less distinct than those in TWA \citep{luh22sc}, likely
because of variable extinction. In addition, since they are much richer
and have more complex structure, Upper Sco and UCL/LCC may not have the
same degree of coevality as a small group like TWA.

Unlike the $G_{\rm BP}-G_{\rm RP}$ CMD, the CMD for
$G-G_{\rm RP}$ contains a large group of stars that
are significantly redder than the TWA sequences for single stars or binaries.
Nearly all of those stars appear in the elevated sequence of unresolved
binaries in the $G_{\rm BP}-G_{\rm RP}$ CMD, which is why the
binary sequence is less populated in $G-G_{\rm RP}$. 
$G$-band fluxes can be underestimated for extended sources \citep{rie21},
which would explain the anomalous colors of those TWA members if they
are marginally resolved binaries.

When they were discovered, the young L dwarfs WISEA J114724.10$-$204021.3
and 2MASS J11193254$-$1137466 were classified as likely members of TWA
\citep{kel15,kel16,sch16}, but more recent astrometry appears
to be inconsistent with membership \citep{bes20}.
Both objects are too faint for detections by Gaia, so I have calculated
their proper motion offsets using the most accurate proper motions that
are available \citep{bes18}, a parallactic distance for 
WISEA J114724.10$-$204021.3 \citep{bes20}, and a photometric distance 
for 2MASS J11193254$-$1137466 \citep{kel15,bes17}. The resulting offsets are 
($\Delta\mu_{\alpha}$, $\Delta\mu_{\delta}$)=$(-9\pm12,-32\pm5)$ and
($-7^{+31}_{-52}, -39^{+14}_{-24}$)~mas~yr$^{-1}$ for
WISEA J114724.10$-$204021.3 and 2MASS J11193254$-$1137466, respectively,
which differ significantly from the measurements for bona fide TWA members
(Figure~\ref{fig:pp}), supporting the results of \citet{bes20}.

\subsection{Spectroscopy of Candidates}
\label{sec:spec}

\begin{figure}
\epsscale{1.4}
\plotone{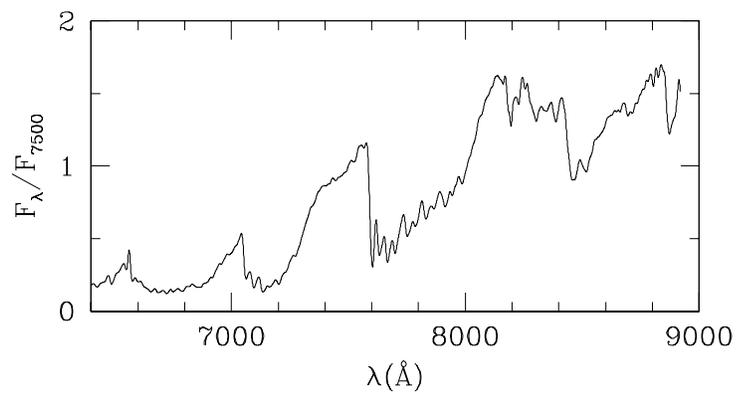}
\caption{
Example of a GMOS spectrum of a candidate member of TWA
(Gaia DR3 5401389770971149568), which is displayed at a resolution of 13~\AA. 
The data used to create this figure are available.
}
\label{fig:spec}
\end{figure}

For candidate members of TWA that lack spectral classifications from
previous studies, I have searched for spectra in the data archives of various
observatories. I found optical spectra of five candidates that were obtained 
with the Fibre-fed Optical Echelle Spectrograph \citep[FEROS;][]{kau99} 
on the MPG/ESO 2.2~m telescope through program 090.C-0200(A),
the Gemini Multi-Object Spectrograph \citep[GMOS;][]{hoo04} at the
Gemini South telescope through program GS-2014A-Q-44,
and the Ultraviolet and Visual Echelle Spectrograph \citep[UVES;][]{dek00}
on the Unit Telescope 2 of the Very Large Telescope
through program 092.C-0203(A). David Rodriguez was the principal 
investigator for each of the three programs.
For eight remaining candidates that lacked archival spectra,
I obtained data with GMOS at the Gemini South telescope.
All of the spectra in question cover a wide range of wavelengths, 
spanning from 6000--9000~\AA\ at a minimum. The data from GMOS, UVES, and
FEROS exhibit $R\sim$4400, 20,000, and 48,000, respectively.
An example of one of the GMOS spectra is shown in Figure~\ref{fig:spec}.
All of the reduced spectra from GMOS are available in an electronic file 
associated with that figure. Pipeline-reduced spectra for UVES and FEROS can
be retrieved from the data archive of the European Southern Observatory.

The 13 candidates with spectra are listed in Table~\ref{tab:spec}.
I have assessed the ages of the targets using Li absorption at 6707~\AA\ and
the Na doublet near 8190~\AA, which is sensitive to surface gravity.
In all of the spectra, the Na doublet is weak enough to be consistent with
a pre-main-sequence star ($\lesssim50$~Myr). Ten of the stars have
sufficiently strong Li absorption (Table~\ref{tab:spec}) to be near the age of 
TWA or younger \citep[e.g.,][]{men08}.
Two stars, which are components of a $2\arcsec$ pair, have upper limits on
the strength of Li that are inconsistent with membership ($<0.1$~\AA),
so they are treated as nonmembers. The final candidate, which is the
coolest one (M6), lacks a useful constraint on Li. For that star,
I take the Na doublet as sufficient evidence of youth for inclusion in
the catalog of members. I have measured spectral types for the candidates
through comparison to field dwarf standards for $<$M5 
\citep{hen94,kir91,kir97} and averages of dwarf and giant standards for 
$\geq$M5 \citep{luh97,luh99}. The resulting classifications range from
M3.5--M6 and are included in Table~\ref{tab:spec}.

\subsection{$UVW$ Velocities}
\label{sec:uvw}

The membership of the TWA candidates selected from Gaia DR3 in 
Section~\ref{sec:ident} can be further constrained via $UVW$ velocities
if they have measurements of radial velocities. I have compiled 
previous measurements of radial velocities for the bona fide members of 
TWA from \citet{gag17} and the additional candidate members.
For most stars that have multiple radial velocity measurements,
I have adopted the velocity with the smallest error.
In the case of Gaia DR3 3567379121431731328, the measurements
span a range of $\sim20$~km~s$^{-1}$, which is much larger than the
individual errors \citep[1--3~km~s$^{-1}$,][Gaia DR2 and DR3]{gag17,rie17b}
and is suggestive of a spectroscopic binary.
For the purpose of calculating the $UVW$ velocity, I have adopted the 
radial velocity from Gaia DR3, which is near the middle of the velocity range.
For TWA~4A, I have used a measurement of the radial velocity for the 
TWA~4A+B system \citep{zun21b}. TWA~23 is a spectroscopic binary that 
exhibits significant variation in its radial velocity measurements 
\citep{shk11,bai12}. \citet{zun21a} calculated the median and standard 
deviation of its available velocities, which I have adopted.
Gaia DR3 provides separate velocities for the components of
the $0\farcs58$ pair TWA~16 while a much more precise measurement
is available for the composite of the system from \citet{zun21a}.
The latter measurement has been adopted for the primary.
Similarly, I have assigned the unresolved radial velocity measurement for
the $0\farcs66$ pair TWA~32 to its primary. The secondary in that system
is the only bona fide member that lacks a radial velocity in my catalog.

I have used the radial velocities in conjunction
with proper motions from Gaia DR3 and parallactic distances based on DR3
parallaxes \citep{bai21} to calculate $UVW$ velocities \citep{joh87}.
The velocity errors were estimated in the manner described by \citet{luh20u}
using the python package {\tt pyia} \citep{pri21}.

\begin{figure}
\epsscale{1.2}
\plotone{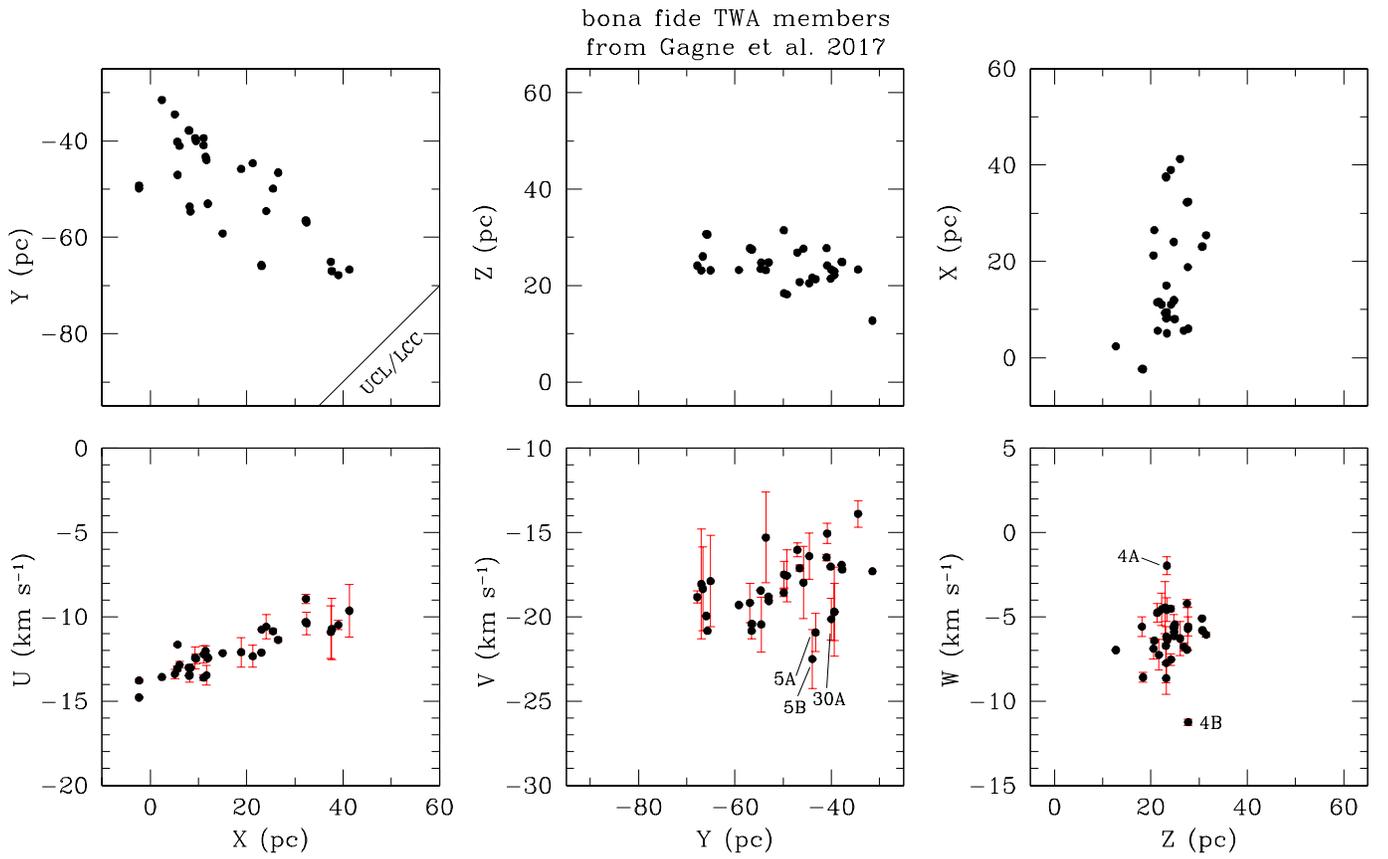}
\caption{
Galactic Cartesian coordinates and $UVW$ velocities for the
bona fide members of TWA compiled by \citet{gag17}. 
The most discrepant measurements of $UVW$ are labeled with the TWA numbers.
}
\label{fig:uvw}
\end{figure}

To illustrate their spatial distribution, I have plotted the bona fide
TWA members from \citet{gag17} in diagrams of $XYZ$ Galactic Cartesian positions
in the top row of Figure~\ref{fig:uvw}.
In the diagram of $Y$ versus $X$,
I have included a boundary that approximates the edge of the UCL/LCC members
identified by \citet{luh22sc}. TWA and UCL/LCC overlap in $Z$, so similar
boundaries are not shown in the other two diagrams of spatial positions.
As mentioned in Section~\ref{sec:ident}, I have excluded the volume 
encompassing UCL/LCC from my survey for TWA members.

In the bottom row of Figure~\ref{fig:uvw}, the measurements of $U$, $V$, and 
$W$ for the bona fide TWA members are plotted versus $X$, $Y$, and $Z$, 
respectively.  Most of the stars are well-clustered in those diagrams.
The outliers are components of binary systems and their binarity likely
accounts for the discrepant velocities. The velocity components of the bona
fide members are correlated with their corresponding spatial dimensions,
which indicates the presence of expansion, as found with the proper motion
offsets in Figure~\ref{fig:pp} (Section~\ref{sec:ident}).
In my survey for new members, I have excluded stars that have $UVW$ velocities
that do not follow the pattern of expansion of the bona fide members, as
discussed in Section~\ref{sec:ident}.

In Figure~\ref{fig:uvw2}, I present diagrams of $XYZ$ and $UVW$ for
the bona fide members and the additional stars that I have adopted as members.
The latter span a larger spatial volume than the former. 
The range of distances among the adopted members is 34--105~pc.
All else being equal, spatial outliers are more likely than
centrally concentrated candidates to be contaminants, so I have 
flagged the outliers in the catalog presented in Table~\ref{tab:mem}.
Ten of the adopted members are absent from the diagrams of $UVW$ since
they lack radial velocity data. One of them, TWA~32B, is 
a secondary in a system that has a measured radial velocity.
Measurements of the radial velocities for the remaining nine stars
would help to better constrain their membership. 
In addition, the available radial velocities for two of the coolest 
members, 2MASS J12474428$-$3816464 and TWA~40, have large
uncertainties (6 and 7~km~s$^{-1}$), so their membership constraints would
benefit from more accurate measurements.
For the 57 adopted members that have measured $UVW$ velocities,
the median velocity is $U, V, W = -12.2, -18.4, -6.1$~km~s$^{-1}$.

The 67 Gaia sources that are adopted as TWA members in this work are presented 
in Table~\ref{tab:mem}, which includes source names from Gaia DR3 and previous
studies; equatorial coordinates, proper motion, parallax, 
renormalized unit weight error \citep[RUWE,][]{lin18}, and photometry
from Gaia DR3; measurements of spectral types and the type adopted in this
work; distance estimate based on the Gaia DR3 parallax \citep{bai21}; 
the adopted radial velocity measurement; the $UVW$ velocities calculated in
this section; the designations and angular separations of the closest
sources within $3\arcsec$ from 2MASS and WISE; flags indicating whether
the Gaia source is the closest match in DR3 for the 2MASS and WISE sources;
photometry from 2MASS and WISE (only for the Gaia source that
is closest to the 2MASS/WISE source); flags indicating whether excesses are
detected in three WISE bands and a disk classification if excess emission
is detected (Section~\ref{sec:disks}); a flag for the bona fide members
from \citet{gag17}; and a flag for spatial outliers (Figure~\ref{fig:uvw2}).
TWA~16, TWA~32, and TWA~39 are binaries with separations of $\lesssim1\arcsec$
in which the components have similar apparent magnitudes from Gaia.
For each system, the spectral classification is based on seeing-limited data
and likely applies to a composite of both components. I have assigned those
spectral types to the primaries in Table~\ref{tab:mem}.

\subsection{Comparison to Previous Studies}

I discuss two stars that are among my adopted TWA members but that have been
rejected in some previous studies: Gaia DR3 3567379121431731328 and TWA~31. 
The former was rejected by \citet{gag17}, likely because of
their measurement of its radial velocity ($-0.1\pm0.8$~km~s$^{-1}$). 
When I adopt the value of $12.29\pm4.15$~km~s$^{-1}$ from Gaia DR3, which
is closer to the middle of the available measurements, the resulting $UVW$
velocity is roughly consistent with membership (Figure~\ref{fig:uvw2}).
The star appears above the single-star sequence for TWA in the CMDs
in Figure~\ref{fig:cmd1}, which is consistent with the fact that it may
be an unresolved binary based on its large spread in measured velocities
and its spectrum \citep{gag17}.  
I have included it in my catalog of adopted members, but a measurement of its
system velocity would be useful for better assessing membership.
The second star, TWA~31, was identified as a possible member of TWA by
\citet{shk11} but was classified as a likely contaminant from LCC by
\citet{gag17}. More recently, \citet{ven19} found that the $UVW$ velocity
of the star based on Gaia DR2 data was consistent with membership in TWA.
I arrive at the same conclusion with the data from Gaia DR3. With a position 
near $X, Y, Z=27,-66, 40$~pc, TWA~31 is not a spatial outlier compared to 
other TWA members and it does not overlap with UCL/LCC (Figure~\ref{fig:uvw2}). 
The strength of its Li absorption indicates an age of $<$20~Myr
\citep{shk11}, which is consistent with TWA membership. The presence
of a circumstellar disk with a high accretion rate serves as additional
evidence of youth \citep{shk11,sch12a,ven19}.
It is much fainter than other TWA members near its spectral type in the
Gaia CMDs (Figure~\ref{fig:cmd1}), which suggests that it is seen
primarily in scattered light at optical wavelengths (e.g., edge-on disk).

\citet{gag17} presented three categories of possible members of TWA.
My catalog includes all of their bona fide members (35 Gaia sources in
24 systems), nine of their 12 high-likelihood candidate members
(TWA~3A and B are counted separately), and five of their 44 candidate members.
In addition, it contains four stars rejected for membership in TWA by that
study (e.g., the aforementioned TWA~31 and Gaia DR3 3567379121431731328)
and five of the nine TWA candidates identified with Gaia DR2 by \citet{gag18d}.
One remaining star, Gaia DR3 5412403269717562240, was classified as a 
member of the $\beta$~Pic moving group by \citet{sch19}. It is a modest outlier 
in $Z$ relative to other TWA members ($Z=5$~pc), but its $UVW$ velocity is
consistent with membership when accounting for the pattern of expansion.
It is also an outlier in $Y$ and $Z$ relative to members of $\beta$~Pic
\citep{shk17}.

\section{Properties of the TWA Stellar Population}
\label{sec:pop}

\subsection{Initial Mass Function}
\label{sec:imf}

\begin{figure}
\epsscale{1.2}
\plotone{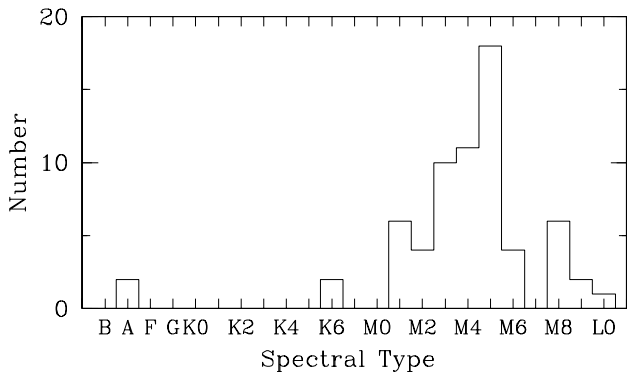}
\caption{
Histogram of spectral types for adopted members of TWA.}
\label{fig:histo}
\end{figure}

Previous catalogs of proposed members of TWA have been used to estimate
the association's IMF \citep{loo11,gag17}.
The same can be done with my new catalog of adopted members.
As in my previous surveys of star-forming regions and young associations,
I have used spectral type as an observational proxy for stellar mass when
characterizing the IMF of TWA. In Figure~\ref{fig:histo}, I have plotted
a histogram of spectral types for the TWA members.
The source catalog from Gaia DR3 has a high level of completeness at
$G\lesssim19$--20 for most of the sky \citep{bou20,fab21}.
At the far side of the association ($\sim100$~pc), $G=19$ corresponds to
a spectral type of $\sim$M8 among members of TWA, so the histogram in
Figure~\ref{fig:histo} should be unbiased for types earlier than that value.
Two likely members that lack Gaia detections are absent from my catalog
and thus do not appear in Figure~\ref{fig:histo} (Section~\ref{sec:ident}),
consisting of TWA~2B \citep[M2--M3.5,][]{web99,her14} and TWA~27B
\citep[mid-to-late L,][]{cha04,moh07,pat10,all13}.
The histogram of spectral types for TWA exhibits a maximum near M5 
($\sim0.15$~$M_\odot$), which resembles the distributions measured
for other nearby young associations \citep[e.g.,][]{luh22sc}.

\subsection{Kinematic Ages}
\label{sec:uvw2}

\begin{figure}
\epsscale{1.2}
\plotone{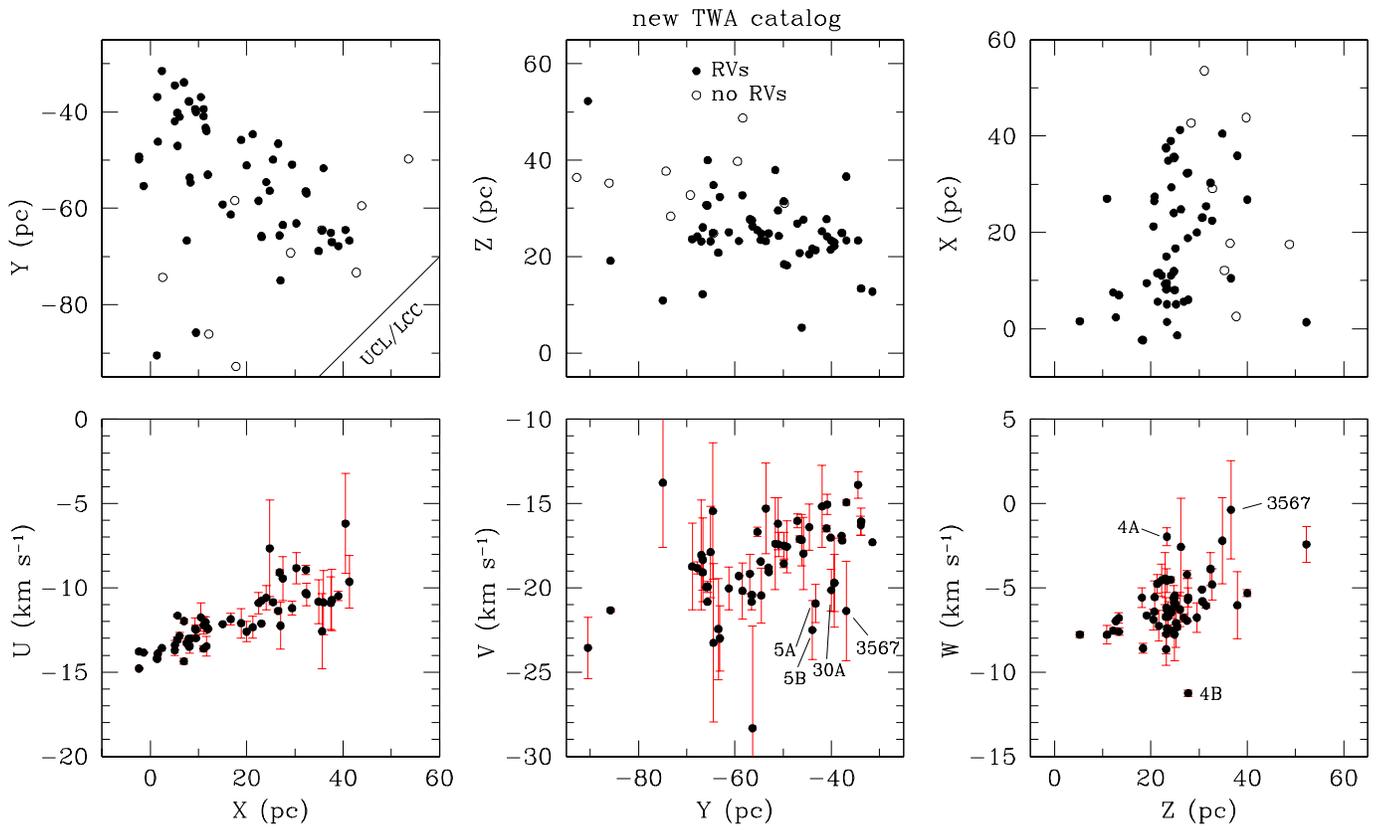}
\caption{
Galactic Cartesian coordinates and $UVW$ velocities for adopted members of TWA
from this work. The most discrepant measurements of $UVW$ are labeled with the 
TWA numbers. The $V$ and $W$ velocities of Gaia DR3 3567379121431731328 are
also marked. Its membership is discussed in Section~\ref{sec:uvw}.
}
\label{fig:uvw2}
\end{figure}

Two forms of kinematic ages can be explored for young unbound associations,
consisting of the expansion age and the traceback age
\citep[e.g.,][]{bla64,bro97,mam14,cru19,zar19,mir20,luh22o,cou23,gal23}.
Previous studies have attempted to derive both ages for TWA
\citep{mak05,mam05,del06,duc14,don16}.
Kinematic age estimates in TWA should benefit significantly from the expanded
size and improved membership classifications in the new TWA census, the 
high precision of the Gaia astrometry, and the availability of radial 
velocity data for most of the members.
For my analysis of the kinematic ages, I have omitted the two members that have 
the largest radial velocity uncertainties, 2MASS J12474428$-$3816464 and 
TWA~40, and multiple systems that have discrepant velocities relative to other 
TWA embers, which consist of TWA~4A/B, TWA~5A/B, and Gaia DR3 
3567379121431731328 (Figure~\ref{fig:uvw2}).

As mentioned in Section~\ref{sec:uvw} and illustrated in Figure~\ref{fig:uvw2},
the space velocities for the TWA members exhibit correlations with spatial
positions, which indicate the presence of expansion.
The rate of expansion can be directly converted to an age under the assumption
that the stars have traveled on linear trajectories.
However, members of an expanding association will eventually experience
acceleration due to the Galactic potential \citep{bla52}.
For an association near an age of 10~Myr, the effect on the expansion
rates in $X$ and $Y$ should be negligible, but the rate in $Z$ could differ
by up to 20--30\% from the value for linear expansion \citep{bro97}.
I have estimated the slopes of the correlations between $UVW$ and $XYZ$
in Figure~\ref{fig:uvw2}, which are the expansion rates, using robust
linear regression with bootstrap sampling, arriving at $0.103\pm0.012$, 
$0.099\pm0.021$, and $0.101\pm0.018$~km~s$^{-1}$~pc$^{-1}$ in $X$, $Y$, and 
$Z$, respectively. In studies of young associations, the $XYZ$ expansion rates
often are inconsistent with a single well-defined value, but the
rates for TWA have fairly small errors and agree surprisingly well,
including the rate in $Z$.
The weighted mean of the three slopes is $0.102\pm0.009$~km~s$^{-1}$~pc$^{-1}$,
which corresponds to an expansion age of 9.6$^{+0.9}_{-0.8}$~Myr.
As done for the 32~Ori association in \citet{luh22o}, I have also
attempted to estimate a traceback age for TWA by identifying the times in
the past when the standard deviations of $X$, $Y$, and $Z$ were each minimized.
The implied ages do not correspond to a single value
and instead range from $\sim$3--8~Myr. It is possible that choosing
a different metric for the size of the association would produce a better
defined traceback age \citep{mir20,cou23,gal23}.

\subsection{Isochronal Age}
\label{sec:iso}

The age of a young association can be estimated from a comparison of
its sequence of low-mass stars in an H-R diagram
to isochrones from theoretical evolutionary models
\citep{bar15,cho16,dot16,fei16}.
Previous studies have applied that method to members of TWA, typically 
producing ages of $\sim8$--10~Myr 
\citep{sta95,sod98,web99,bar06,wei13,bel15,her15,don16}.
As with the kinematic ages, it should be possible to improve the
estimate of the isochronal age of TWA with the new census.

\begin{figure}
\epsscale{1.1}
\plotone{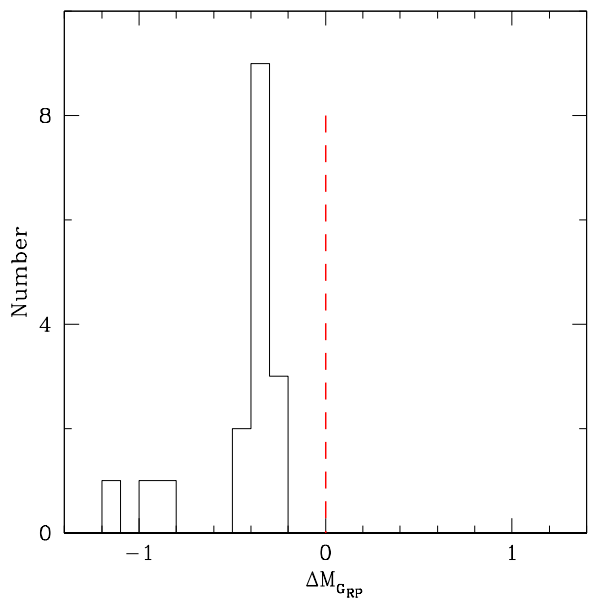}
\caption{
Histogram of offsets in $M_{G_{\rm RP}}$ from the median CMD sequence for
UCL/LCC for low-mass stars in TWA (Figure~\ref{fig:cmd1}).
Negative values correspond to brighter magnitudes and younger ages.}
\label{fig:ages}
\end{figure}

To derive an isochronal age for TWA, I have followed the procedure 
applied to associations near the Taurus star-forming region in \citet{luh23}.
I have selected TWA members that have precise parallaxes
($\sigma_{\pi}<0.1$~mas) and photometry ($\sigma_{BP}<0.1$, $\sigma_{RP}<0.1$)
and have colors indicative of low-mass stars ($G_{\rm BP}-G_{\rm RP}=1.4$--2.8).
Disk-bearing stars are excluded to avoid contamination of the
photometry by disk-related emission.
Within that color range, all members satisfy the criteria for photometric
errors and only four of the diskless stars are omitted because
of parallax errors. All of the latter are companions that have poor
astrometric fits according to their large values of RUWE.
The resulting sample contains 17 sources.
I have calculated their offsets in $M_{G_{\rm RP}}$ from a fit to the median
of the sequence for UCL/LCC \citep{luh23}. I have assumed that the stars
have no extinction since they are located within the Local Bubble.
A histogram of the $M_{G_{\rm RP}}$ offsets is shown in Figure~\ref{fig:ages}.
I have calculated the median and the median absolute deviation (MAD) of the
offsets, arriving at $-$0.36 and 0.07, respectively. 
If the criteria for parallax error or disks are eliminated, the median
offset is nearly unchanged, but the MAD increases because most of the
added stars are elevated above the single-star sequence (i.e., likely
unresolved binaries). This is a reflection of the fact that the value of MAD 
is dominated by the intrinsic width of the TWA sequence produced by single
and binary stars rather than measurement errors.

I have converted the median offset for the sample of 17 stars to an
age by assuming that UCL/LCC ($\Delta$M=0) has an age of 20~Myr
\citep{luh22sc} and that $\Delta$log~L/$\Delta$log~age$=-0.6$, as done in
\citet{luh23}. The age of 20~Myr for UCL/LCC is tied to the lithium depletion
boundary (LDB) age for the $\beta$~Pic moving group \citep{bin16}, as 
discussed in \citet{luh22sc}. The resulting isochronal age for TWA is
$11.4^{+1.3}_{-1.2}$~Myr. Once again, that error is determined primarily
by the gap between the single- and double-star sequences and the relative
numbers of singles and binaries in the sample. 
In addition, the derived age may be subject to systematic errors arising
from the adopted age for UCL/LCC and the conversion from
$M_{G_{\rm RP}}$ offsets to relative ages.

As an alternative to the above approach, an age for TWA can be estimated by
directly fitting its single-star sequence of low-mass stars in $M_{G_{\rm RP}}$ 
versus $G_{\rm BP}-G_{\rm RP}$ with model ishochrones.
For instance, the isochrones of \citet{bar15} imply an age of
$\sim6$~Myr while the models of \citet{dot16} and \citet{cho16}
imply an age of $\sim4$~Myr.
Models that account for the inhibition of convection by magnetic fields
tend to produce older ages that are in better
agreement with ages derived from the LDB
and the mass-radius relationship for young low-mass stars
\citep{kra15,dav16a,dav19b,fei16,mac17,bin22}.
However, in the Gaia CMD, new versions of the magnetic isochrones from
\citet{fei16} provided by G. Feiden (private communication)
are much less parallel to the TWA single-star sequence than the non-magnetic
isochrones, implying ages of $\sim$25 and 11~Myr at 
$G_{\rm BP}-G_{\rm RP}=1.6$ and 3, respectively.
Given these complications with the fitting of both non-magnetic and
magnetic isochrones, I prefer the procedure followed earlier in this section
in which empirical isochrones are compared and some are dated with the
LDB \citep{her15}.

From this study, the most reliable age estimates for TWA appear to be the
expansion age of 9.6$^{+0.9}_{-0.8}$~Myr and the isochronal age of 
$\sim$11.4~Myr. Based on these values, I adopt an age of 10$\pm$2~Myr for TWA.

\subsection{Circumstellar Disks}
\label{sec:disks}

I have used mid-IR photometry from WISE to search for evidence of disks
among the adopted members of TWA. The 67 objects in my catalog 
have 61 matching sources from WISE.  If a close pair of Gaia sources has the
same WISE source as their closest match, the WISE designation appears in both
of their entries in Table~\ref{tab:mem}, but the disk measurements from this 
section are listed only for the candidate that is closest to the WISE source.
The AllWISE Atlas images of the WISE sources have been visually inspected to
check for detections that are false or unreliable, which are marked by a
flag in Table~\ref{tab:mem}.

\begin{figure}
\epsscale{1.2}
\plotone{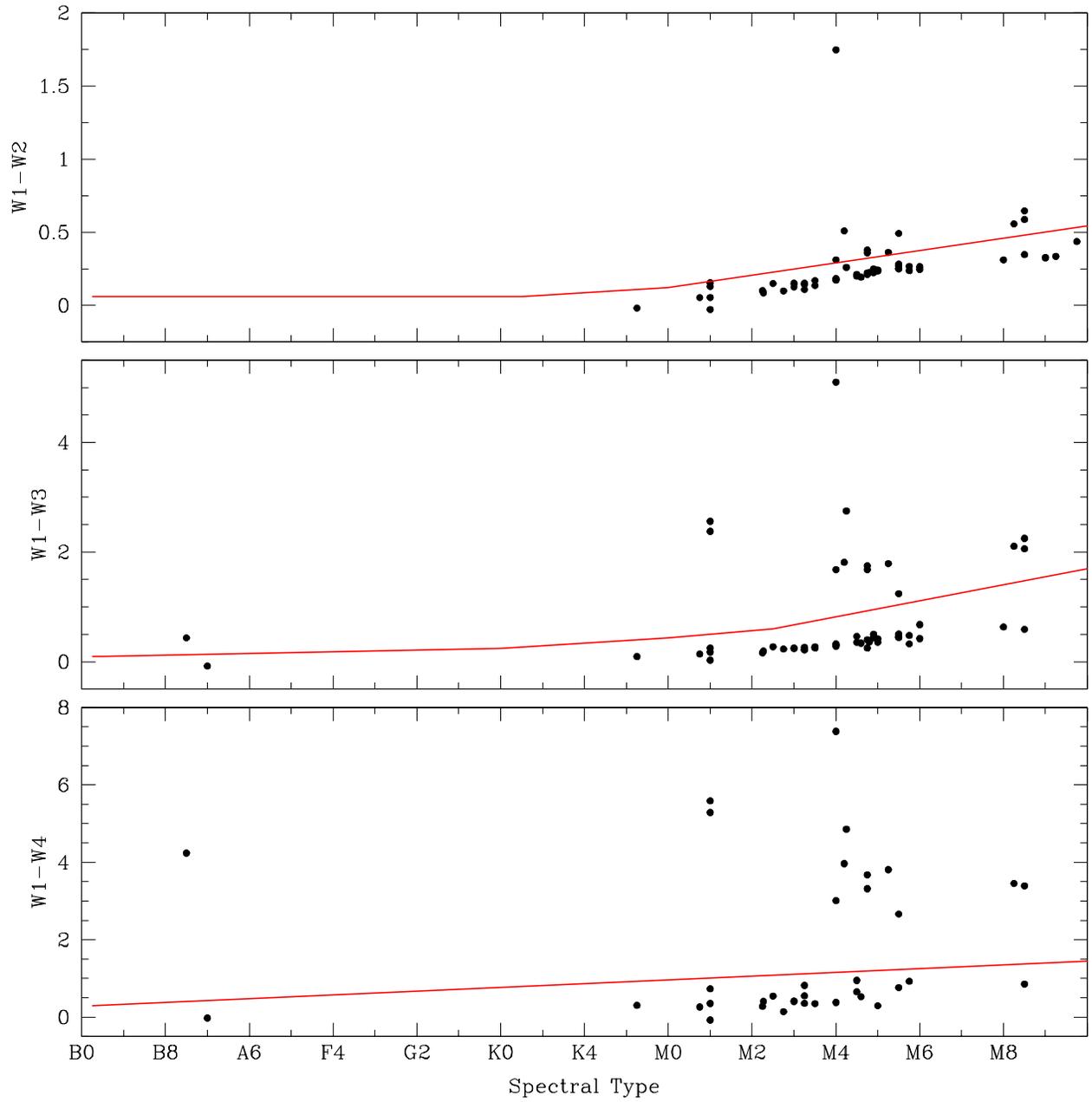}
\caption{ 
IR colors versus spectral type for adopted members of TWA. 
In each diagram, the tight sequence of blue colors corresponds to stellar
photospheres. The thresholds used for identifying color excesses from
disks are indicated \citep[red solid lines,][]{luh22disks}.
}
\label{fig:exc1}
\end{figure}

As done in my previous surveys of young associations 
\citep[e.g.,][]{luh22disks}, I have used W1$-$W2, W1$-$W3, and W1$-$W4 to 
detect excess emission from disks among the 61 WISE sources in TWA.
Those colors are plotted versus spectral type in Figure~\ref{fig:exc1}.
I have omitted W2 data at W2$<$6 since they are subject to significant
systematic errors \citep{cut12b}.
In each of the three colors, some sources are found in a blue sequence that
corresponds to stellar photospheres while others have redder colors
that indicate the presence of IR excess emission from circumstellar dust.
In Figure~\ref{fig:exc1}, I have marked the threshold for each color 
that was used by \citet{luh22disks} for identifying color excesses.
Flags indicating the presence or absence of excesses in W2, W3, and W4
are included in Table~\ref{tab:mem}.

For two of the TWA members that lack detections in W4, TWA~8B and TWA~26,
measurements in a similar band at 24~\micron\ (denoted as [24]) with the 
Multiband Imaging Photometer for Spitzer \citep[MIPS;][]{rie04,wer04} 
have been reported in previous studies \citep{ria06b,luh10,sch12a}.
I have adopted the MIPS data as a proxy for W4 when
calculating W1$-$W4 and determining the excess flag for W4.
Because of the proximity to its primary ($8\arcsec$), HR~4796B appears
to have unreliable photometry in W2, W3, and W4. As a result, it is the
only WISE source in my catalog that is not assessed for excess emission
in any of those bands.

\begin{figure}
\epsscale{1.4}
\plotone{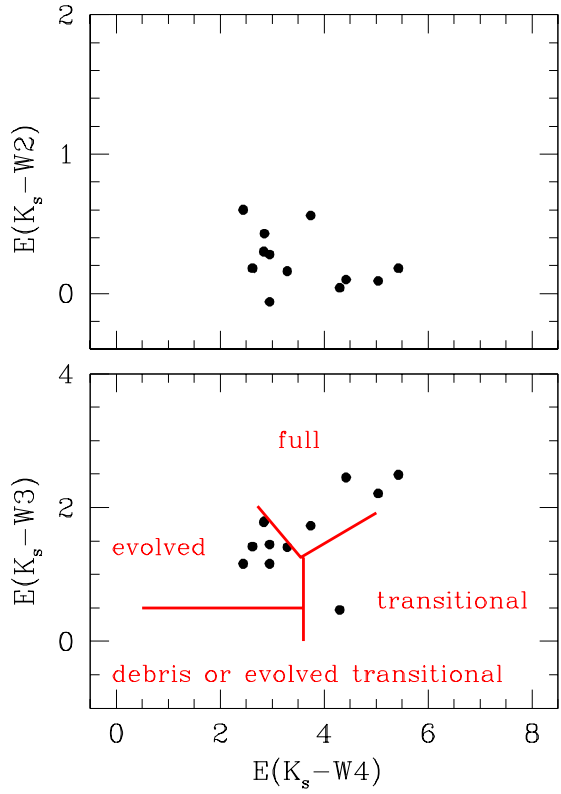}
\caption{
IR color excesses for adopted members of TWA. The boundaries used for assigning
disk classes are shown in the bottom diagram (red solid lines).
}
\label{fig:exc2}
\end{figure}

The evolutionary stages of the detected disks have been classified from among
the following options: full disk, transitional disk, evolved disk, evolved
transitional disk, and debris disk \citep{ken05,rie05,her07,luh10,esp12}.
The classes are estimated based on the sizes of the excesses in
$K_s-$W3 and $K_s-$W4 \citep{luh12u,esp14,esp18}.
The color excesses, E($K_s-$W3) and E($K_s-$W4), are calculated by
subtracting the expected photospheric color for a given spectral type
\citep{luh22sc}. The resulting excesses are plotted in Figure~\ref{fig:exc2}
with the criteria for the disk classes \citep{esp18}.  The values of 
E($K_s-$W2) are included as well to illustrate sizes of the excesses in W2. 
Sources that lack excesses in any of the WISE bands are excluded
from Figure~\ref{fig:exc2} and are designated as class~III \citep{lw84,lad87}.
The excesses for TWA~30B are large enough to place it beyond the limits
of Figure~\ref{fig:exc2}. Gaia DR3 6152893526035165312 has excesses in W2 
and W3 but is not detected in W4, so it is absent from those diagrams.
The excesses in Figure~\ref{fig:exc2} imply that HR~4796 has a transitional
disk and that TW~Hya and HD~98800B have full disks, but more detailed
measurements of their spectral energy distributions indicate that
HR~4796 has a debris disk and the latter two stars have transitional disks
\citep{jur91,aug99,cal02,uch04,fur07}, which are the classes adopted in 
Table~\ref{tab:mem}.

Fourteen of the WISE sources exhibit IR excess emission, all of which have had
disks detected in previous work
\citep{ruc83,del89,jur91,gre92,zuc93,moh03,ste04,uch04,low05,ria06b,mor08,reb08,ria08,her09,loo10a,loo10b,shk11,sch12a,sch12b,rod15,bou16,bin17}.
They consist of four full, seven evolved, two transitional, and one
debris disk. Previous studies have reported a modest excess in [24] for 
TWA~7 based on $K_s-[24]$ \citep{low05,reb08,sch12a,bin17}, but I do not find 
an excess in either [24] or W4 relative to W1. The previous detections 
of an excess in [24] may have been caused by variability between the 
observations in $K$ and the mid-IR bands. Similarly, \citet{reb08} 
identified an excess in [24] for TWA~8B using $K_s-[24]$, whereas I find that 
W1$-$[24] is consistent with the expected photospheric value.

Among WISE sources in TWA that have been assessed for IR excesses and that
have spectral types of $\leq$M6, the fraction that have full, transitional,
or evolved disks is 10/52=$0.19^{+0.08}_{-0.06}$.
That value is consistent with the disk fraction for the same range of 
types in Upper Sco \citep[$0.19\pm0.01$,][]{luh22disks},
which is roughly coeval with TWA \citep[Section~\ref{sec:iso},][]{luh20u}.

\section{Conclusions}

I have performed a survey for members of the TW Hya association using
high-precision photometry and astrometry from Gaia DR3 and ground-based
spectroscopy. I have used the new catalog of adopted members to
characterize the IMF and the age of the association and to identify and
classify its circumstellar disks. The results are summarized as follows:

\begin{enumerate}

\item
\citet{gag17} compiled a sample of bona fide members of TWA, consisting
of 24 systems that are resolved into 35 sources in Gaia DR3.
I have used the kinematics and photometry of those members to guide a search
for additional members with Gaia DR3.
For the resulting candidates, I have checked whether membership is supported
by $UVW$ velocities (when radial velocities are available) and spectroscopic
diagnostics of age. All of the candidates have the latter
from previous studies or new spectra analyzed in this work.
In the end, my catalog of adopted members contains 67 Gaia sources
in 55 systems. In addition to the bona fide members, the catalog includes
nine high-likelihood candidates (eight systems) and five candidates
from \citet{gag17} and five candidates from \citet{gag18d}.
The remaining 13 adopted members have not been previously classified
as candidates. Although all objects in the catalog have good evidence
of membership from Gaia data and spectroscopy, measurements of radial
velocities for the nine systems that lack such data would be useful 
for further assessing their membership.

\item
The histogram of spectral types for the adopted members of TWA
peaks near M5 ($\sim0.15$~$M_\odot$), indicating that the characteristic
mass of the IMF in TWA is similar to the values in other nearby associations
and star-forming regions.

\item
For the TWA members that have measured radial velocities,
$U$, $V$, and $W$ are positively correlated with $X$, $Y$, and $Z$,
respectively, indicating the presence of expansion.
The slopes of the three correlations are consistent with each other
and their weighted mean corresponds to an expansion age of
9.6$^{+0.9}_{-0.8}$~Myr.

\item
In a CMD constructed from Gaia data, the members of TWA exhibit
well-defined sequences of single stars and unresolved binary stars.
TWA is one of the youngest populations (if not the youngest) in which
the two sequences have been resolved.
The combined sequence of low-mass stars is $0.36\pm0.07$~mag brighter
than the median sequence for UCL/LCC.
If one adopts an age of 20~Myr for the latter and assumes that
young low-mass stars fade at a rate given by
$\Delta$log~L/$\Delta$log~age$=-0.6$, the offset for TWA from UCL/LCC
corresponds to an age of $11.4^{+1.3}_{-1.2}$~Myr. Based on that value and 
the expansion age, I adopt an age of 10$\pm$2~Myr for TWA.

\item
I have used mid-IR photometry from WISE to identify TWA members
that exhibit IR excesses from disks. The evolutionary stages of the 
detected disks have been classified using the sizes of the excesses.
Fourteen members have IR excesses, all of which have had disks reported
in previous studies.
Among WISE sources in TWA that have been assessed for IR excesses and that
have spectral types of $\leq$M6, the fraction that have full, transitional,
or evolved disks is 10/52=$0.19^{+0.08}_{-0.06}$.
That value is consistent with the disk fraction for the same range of 
types in Upper Sco \citep[$0.19\pm0.01$,][]{luh22disks},
which is roughly coeval with TWA.

\end{enumerate}

\acknowledgements

This work used data from the European Space Agency 
mission Gaia (\url{https://www.cosmos.esa.int/gaia}), processed by
the Gaia Data Processing and Analysis Consortium (DPAC,
\url{https://www.cosmos.esa.int/web/gaia/dpac/consortium}). Funding
for the DPAC has been provided by national institutions, in particular
the institutions participating in the Gaia Multilateral Agreement. The Gemini
data were obtained through programs GS-2014A-Q-44 and GS-2023A-FT-20.
Gemini Observatory is a program of NSF's NOIRLab, which is managed by the
Association of Universities for Research in Astronomy (AURA) under a
cooperative agreement with the National Science Foundation on behalf of the
Gemini Observatory partnership: the National Science Foundation (United States),
National Research Council (Canada), Agencia Nacional de Investigaci\'{o}n y
Desarrollo (Chile), Ministerio de Ciencia, Tecnolog\'{i}a e Innovaci\'{o}n
(Argentina), Minist\'{e}rio da Ci\^{e}ncia, Tecnologia, Inova\c{c}\~{o}es e
Comunica\c{c}\~{o}es (Brazil), and Korea Astronomy and Space Science Institute
(Republic of Korea). 2MASS is a joint project of the University of 
Massachusetts and IPAC at Caltech, funded by NASA and the NSF.
WISE is a joint project of the University of California, Los Angeles,
and the JPL/Caltech, funded by NASA. This work used data from the 
NASA/IPAC Infrared Science Archive, operated by JPL under contract
with NASA, and the VizieR catalog access tool and the SIMBAD database, 
both operated at CDS, Strasbourg, France.
The Center for Exoplanets and Habitable Worlds is supported by the
Pennsylvania State University, the Eberly College of Science, and the
Pennsylvania Space Grant Consortium.

\clearpage

\LongTables

\begin{deluxetable}{lllll}
\tabletypesize{\scriptsize}
\tablewidth{0pt}
\tablecaption{New Spectral Classifications for Candidate Members of TWA\label{tab:spec}}
\tablehead{ 
\colhead{Gaia DR3} &
\colhead{Spectral Type} &
\colhead{$W_{\lambda}$(Li)\tablenotemark{a}} &
\colhead{Instrument} &
\colhead{Date}\\
\colhead{} &
\colhead{} &
\colhead{(\AA)} &
\colhead{} &
\colhead{}}
\startdata
5414158429569765632 & M3.5 & 0.48 & FEROS & 2013 Feb 17 \\
5397574190745629312 & M4 & 0.55 & FEROS & 2013 Feb 17 \\
6183591791897683584 & M4 & 0.47 & FEROS & 2013 Feb 17 \\
6133420114251217664 & M5.5 & 0.45 & UVES & 2013 Dec 27 \\
6145303429765430784 & M4.9 & 0.80 & GMOS & 2014 Feb 22 \\
3493814268751183744 & M5.75 & 0.64 & GMOS & 2023 Mar 5 \\
5348165127505382400 & M4.75 & 0.60 & GMOS & 2023 Mar 5 \\
5401389770971149568 & M6    & \nodata & GMOS & 2023 Mar 5 \\
6114656192408518784 & M4.6  & 0.58 & GMOS & 2023 Mar 5 \\
6143632653128880896 & M4.9  & 0.72 & GMOS & 2023 Mar 5 \\
6143984423832713856 & M4.25 & $<$0.1 & GMOS & 2023 Mar 5 \\
6143984428129994624 & M4.6  & $<$0.1 & GMOS & 2023 Mar 5 \\
6179256348830614784 & M4.75 & 0.65 & GMOS & 2023 Mar 5 
\enddata
\tablenotetext{a}{Typical uncertainties are 0.05~\AA.}
\end{deluxetable}

\clearpage

\begin{deluxetable}{ll}
\tabletypesize{\scriptsize}
\tablewidth{0pt}
\tablecaption{Adopted Members of TWA\label{tab:mem}}
\tablehead{
\colhead{Column Label} &
\colhead{Description}}
\startdata
GaiaDR3 & Gaia DR3 source name \\
Name & Other source name \\
RAdeg & Gaia DR3 right ascension (ICRS at Epoch 2016.0)\\
DEdeg & Gaia DR3 declination (ICRS at Epoch 2016.0)\\
SpType & Spectral type \\
r\_SpType & Spectral type reference\tablenotemark{a} \\
Adopt & Adopted spectral type \\
pmRA & Gaia DR3 proper motion in right ascension\\
e\_pmRA & Error in pmRA \\
pmDec & Gaia DR3 proper motion in declination\\
e\_pmDec & Error in pmDec \\
plx & Gaia DR3 parallax\\
e\_plx & Error in plx \\
rmedgeo & Median of geometric distance posterior \citep{bai21}\\
rlogeo & 16th percentile of geometric distance posterior \citep{bai21}\\
rhigeo & 84th percentile of geometric distance posterior \citep{bai21}\\
RVel & Radial velocity \\
e\_RVel & Error in RVel \\
r\_RVel & Radial velocity reference\tablenotemark{b} \\
U & $U$ component of space velocity \\
e\_U & Error in U \\
V & $V$ component of space velocity \\
e\_V & Error in V \\
W & $W$ component of space velocity \\
e\_W & Error in W \\
Gmag & Gaia DR3 $G$ magnitude\\
e\_Gmag & Error in Gmag \\
GBPmag & Gaia DR3 $G_{\rm BP}$ magnitude\\
e\_GBPmag & Error in GBPmag \\
GRPmag & Gaia DR3 $G_{\rm RP}$ magnitude\\
e\_GRPmag & Error in GRPmag \\
RUWE & Gaia DR3 renormalized unit weight error\\
2m & Closest 2MASS source within $3\arcsec$ \\
2msep & Angular separation between Gaia DR3 (epoch 2000) and 2MASS \\
2mclosest & Is this Gaia source the closest match for the 2MASS source? \\
wise & Closest WISE source within $3\arcsec$ \\
wisesep & Angular separation between Gaia DR3 (epoch 2010.5) and WISE \\
wiseclosest & Is this Gaia source the closest match for the WISE source?\\
Jmag & 2MASS $J$ magnitude \\
e\_Jmag & Error in Jmag \\
Hmag & 2MASS $H$ magnitude \\
e\_Hmag & Error in Hmag \\
Ksmag & 2MASS $K_s$ magnitude \\
e\_Ksmag & Error in Ksmag \\
W1mag & WISE W1 magnitude \\
e\_W1mag & Error in W1mag \\
W2mag & WISE W2 magnitude \\
e\_W2mag & Error in W2mag \\
f\_W2mag & Flag on W2mag\tablenotemark{c} \\
W3mag & WISE W3 magnitude \\
e\_W3mag & Error in W3mag \\
f\_W3mag & Flag on W3mag\tablenotemark{c} \\
W4mag & WISE W4 magnitude \\
e\_W4mag & Error in W4mag \\
f\_W4mag & Flag on W4mag\tablenotemark{c} \\
ExcW2 & Excess present in W2? \\
ExcW3 & Excess present in W3? \\
ExcW4 & Excess present in W4?\tablenotemark{d} \\
DiskType & Disk type \\
bfmem & Bona fide member of TWA from \citet{gag17} \\
outlier & Spatial outlier 
\enddata
\tablenotetext{a}{
(1) \citet{sch19};
(2) \citet{rie14};
(3) \citet{rie17b};
(4) this work;
(5) \citet{rod11};
(6) \citet{ria06};
(7) \citet{sch12b};
(8) \citet{gag15c};
(9) \citet{mur15};
(10) \citet{luh17};
(11) \citet{web99};
(12) \citet{tor06};
(13) \citet{man13};
(14) \citet{pec13};
(15) \citet{her14};
(16) \citet{gag17};
(17) \citet{her78};
(18) \citet{del89};
(19) \citet{tor00};
(20) \citet{vac11};
(21) \citet{ven19};
(22) \citet{sch05};
(23) \citet{all13};
(24) \citet{loo11};
(25) \citet{hou82};
(26) \citet{can93};
(27) \citet{bid88};
(28) \citet{abt95};
(29) \citet{shk11};
(30) \citet{ste99};
(31) \citet{hou88};
(32) \citet{haw96};
(33) \citet{sod96};
(34) \citet{neu00};
(35) \citet{bon14};
(36) \citet{bin20};
(37) \citet{loo10b};
(38) \citet{loo10a};
(39) \citet{whi04};
(40) \citet{giz02};
(41) \citet{loo07};
(42) \citet{rei08};
(43) \citet{bow19};
(44) \citet{zuc04};
(45) \citet{her09};
(46) \citet{gag14b};
(47) \citet{rei03};
(48) \citet{zuc01};
(49) \citet{kas08};
(50) \citet{sta95};
(51) \citet{man14}.}
\tablenotetext{b}{
(1) \citet{sch19};
(2) Gaia DR3;
(3) \citet{bud18};
(4) \citet{gag17};
(5) \citet{bai12};
(6) \citet{mur15};
(7) \citet{fou18};
(8) \citet{zun21a};
(9) \citet{shk11};
(10) \citet{zun21b};
(11) \citet{tor03};
(12) \citet{moh03};
(13) \citet{loo10b};
(14) \citet{loo10a};
(15) \citet{ven19};
(16) \citet{rei03};
(17) \citet{ell14};
(18) \citet{kid19}.}
\tablenotetext{c}{nodet = nondetection; false = detection from
AllWISE appears to be false or unreliable based on visual inspection.}
\tablenotetext{d}{Excess classifications in W4 for TWA~8B and TWA~26
are based on 24~\micron\ photometry from the Spitzer Space Telescope
\citep{luh10}.}
\tablecomments{
The table is available in its entirety in machine-readable form.}
\end{deluxetable}

\clearpage

\end{document}